# QUANTUM STATISTICAL PHYSICS:

# A NEW APPROACH

## U. F. Edgal[*,†,‡] and D. L. Huber[‡]

*School of Technology, North Carolina A&T State University*

*Greensboro, North Carolina, 27411*

*and*

*Department of Physics, University of Wisconsin-Madison*

*Madison, Wisconsin, 53706*

---

[*] To whom correspondence should be addressed. School of Technology, North Carolina A&T State University, 1601 E. Market Street, Greensboro, NC 27411. Phone: (336)334-7718. Fax: (336)334-7546. Email: ufedgal@ncat.edu.
[†] North Carolina A&T State University
[‡] University of Wisconsin-Madison

# ABSTRACT

The new scheme employed (throughout the "thermodynamic phase space"), in the statistical thermodynamic investigation of classical systems, is extended to quantum systems. Quantum Nearest Neighbor Probability Density Functions (QNNPDF's) are formulated (in a manner analogous to the classical case) to provide a new quantum approach for describing structure at the microscopic level, as well as characterize the thermodynamic properties of material systems. A major point of this paper is that it relates the free energy of an assembly of interacting particles to QNNPDF's. Also. the methods of this paper reduces to a great extent, the degree of difficulty of the original equilibrium quantum statistical thermodynamic problem without compromising the accuracy of results. Application to the simple case of dilute, weakly degenerate gases has been outlined..

# I.   INTRODUCTION

The new mathematical framework employed in the determination of the exact free energy of pure classical systems [1] (with arbitrary many-body interactions, throughout the thermodynamic phase space), which was recently extended to classical mixed systems[2] is further extended to quantum systems in the present paper. We begin the investigation using the familiar method of re-expressing the trace of the Boltzmann operator (sum of states) as a multidimensional integral over a portion or all of classical phase space. This therefore makes possible the use of the concepts of generalized order, the reduced one particle phase space, and the nearest neighbor probability density function (NNPDF), all of which were found to be essential ingredients in the formulation of the exact scheme for classical systems. In the absence of these ingredients, earlier investigations using the above approach have usually employed approximate methods such as the cluster expansion scheme, leading to results applicable only under weakly degenerate conditions. Due to extreme inherent difficulties in quantum systems (unlike the classical case), other methods of approach have had to employ a variety of other kinds of approximation techniques and model systems ($C^*$ algebra, para-statistics, fermi liquid theory, renormalization group, Bosonization, etc.) providing varying degrees of success and insights. The reader is referred to a vast literature.[3] A major purpose of the present paper is to show that the single mathematical framework already enunciated in previous papers[1,2] for pure and mixed systems is not restricted to classical systems alone, but may be employed for all systems (classical and quantum). Numerical aspects of the new scheme have however not yet been completed, and thus remain of top priority for subsequent investigations.

We restrict the current investigation to the case of single component quantum systems, for which we employ (without loss of generality) the coordinate representation. The sum of states written in terms of a multi-dimensional integral for a system of N identical particles interacting via some many-body interaction potential may therefore be given as:

$$\widetilde{Z}(N,V) = \frac{1}{N!\sigma^N} \int ... \int W_N dX_1''...dX_N''$$

(1)

V is volume of the system, $\sigma$ is some constant whose dimension makes $\widetilde{Z}(N,V)$ dimensionless, and $X_1'',...,X_N''$ specify a given micro-state in classical phase space. For simplicity, internal degrees of freedom are not rigorously treated in this paper, providing a strictly non-relativistic treatment. (Spin and other internal degrees of freedom may however still be introduced in the usual manner). In the simplest case where the portion of phase space over which the integral is specified is configuration space, the phase space distribution is the "Slater Sum" written as:

$$W_N(X_1'',...,X_N'') = N!\lambda^{3N} \sum_\alpha \psi_\alpha *(X_1'',...,X_N'')e^{-\beta H_N}\psi_\alpha(X_1'',...,X_N'')$$

(2)

where $\sigma = \lambda^3$, $\{\psi_\alpha\}$ is a complete set of orthonormal wave functions symmetrized or anti-symmetrized (according to whether we have Bosons or Fermions). $W_N$ is the diagonal element of the Boltzmann's operator in coordinate representation. $\beta = 1/KT$ where K is Boltzmann's constant, and T is temperature; $H_N$ is the Hamiltonian operator of the system of N identical particles, and the asterisk * implies complex conjugation. $\lambda$ is the thermal wavelength $\left(\dfrac{h^2}{2\pi mKT}\right)^{\frac{1}{2}}$ where h is Planck's constant, and m is mass of a particle. $X_1'', ..., X_N''$ refer to the translational coordinates of particles.

The product property [4] for the Slater Sum, is well known, and may be stated as follows: $W_N \approx W_n W_{N-n}$, where particle coordinates are separated into two groups, and those used to compute $W_n$ are distinct from those used to compute $W_{N-n}$. Also, any two coordinates $X_i'', X_j''$ belonging to different groups must satisfy the separation condition $r_0, \lambda << \left| X_i'' - X_j'' \right|$. ($r_0$ is the effective inter-particle interaction range). The approximate equality in the product property suggests $W_N$ and $W_n W_{N-n}$ differ by at most a factor not much different than unity. Extensive discussions of this property are available in the literature (to which we refer the reader), and shall not be investigated at this time. We remark however, that in the developments in this paper, the two groups of particle coordinates of interest involve the n nearest neighbors of the "origin" (said to be randomly situated within volume V) on the one hand, and the rest of the (N-n) particles on the other. Because the "surface-to-volume ratio" tends to zero as volume increases, we can always find some n which is finite (though large enough), such that the separation condition stated above may hold for an overwhelming set of coordinate pairs. This should be true except for a relatively small number of coordinate pairs involving particles close to the boundary separating the volume containing the n nearest neighbors and the rest of the system particles. The effect of the few pairs of coordinates become inconsequential when n is large enough. We get some sense of how large n may need to be (in realistic systems) by the following calculations. For an electron mass at $10^{o}$K, we have that $\lambda \sim$ 80Å; and for a neutron mass at $1^{o}$K, $\lambda \sim 10$Å. Hence, we find that for very light particles, and at the lowest temperatures, $\lambda$ is at most $\sim$ 10 lattice constants (for typical solids). Hence a small surface to volume ratio (and thus the product property as stated above) becomes valid at the lowest temperatures, and highest (solid state) densities, when the number of nearest neighbor particles, n, are situated within a volume with a characteristic length typically of the order of several tens of lattice constants. Thus, this amounts to n being probably at most a few thousand or so. For more massive particles, and at moderate temperatures, the required size of n should be considerably smaller.

For the purposes of the present paper, only the following much milder set of statements of the product property are needed, and thus employed: In the first mild statement, we note that $W_N$ may be written rigorously as $W_N = W_n W_{N-n} F_n$ (where $F_n$ is some number). Hence if we may write $W_n = e^{w_n}$, $W_{N-n} = e^{w_{N-n}}$, $F_n = e^{f_n}$, we need $\left| w_n \right|$, $\left| w_{N-n} \right| \gg \left| f_n \right|$ for n, N-n $\gg$ 1. The earlier stated product property assumes $f_n \approx$ 0, implying a more stringent statement. Clearly, the first mild statement allows the possibility of $W_N$ and the product $W_n W_{N-n}$ being quite disparate from one another, as $e^{f_n}$ may be much smaller or much larger than unity. In the second mild statement, we need $f_n - f_{n-m} \rightarrow 0$ as n gets larger (n > m, and m is fixed). Hence we may write:

$$\frac{W_N}{W_{N-m}} = \frac{W_{N-n} W_n F_n}{W_{N-n} W_{n-m} F_{n-m}} \approx \frac{W_n}{W_{n-m}} = e^{w_n - w_{n-m}} \quad \text{for n sufficiently large (n > m).}$$

The provision for n sufficiently large, makes this statement considerably less stringent than the earlier stated product property where $f_n - f_{n-m}$ is always $\approx$ zero for all allowed n. These milder statements of the product property have a logarithmic nature, and thus impose a far less stringent requirement on how large n may need to be. Hence we expect that n might be considerably less than a few thousands at the highest (solid state) densities and lowest temperatures (slightly removed from $0^o$K). How large n may need to be, to obtain accurate results, may ultimately be determined by the Monte Carlo method where the accuracy of the product property may be investigated progressively as N increases from some small value. In the very close neighborhood of zero temperature (where $\lambda \to \infty$) however, we expect we would need a prohibitively large n; implying our scheme may not be used right down to zero temperature. The Slater Sum product property, and its milder forms as presented above, shall be more extensively addressed in a subsequent paper. It should be emphasized that the product property, and thus our current analysis may require some revision (as should also the classical case) in systems that possess such peculiarities as strong density fluctuations (as in phase transition regions), including phase correlations between particles that extend over macroscopic distances (as in superfluids and superconductors with macroscopic quantum coherence). It is however interesting to note that the use of the product property (which has a rigorous foundation), may allow us to avert the phase (or sign) problem usually encountered in calculations involving fermion systems.

In section II of this paper, the concepts of generalized order and the reduced one particle phase space are employed identically as in the classical case[1,2] to develop the statistical thermodynamic formalism for quantum systems. The notion of the Quantum analog of the classical system's Nearest Neighbor Probability Density Function [5] (termed the Quantum NNPDF or QNNPDF) is also introduced in this section. In section III, further exposition on QNNPDF's is briefly summarized. Application to a weakly degenerate quantum system is given in section IV where the second quantum virial coefficient is derived. Broad statements on extensions of the above scheme are given in section V.

# II. STATISTICAL THERMODYNAMIC FORMALISM FOR QUANTUM SYSTEMS

Proceeding as in the classical case, we have that the translational coordinates $X_1'', ..., X_N''$ may be ordered in the sense of one of the stated versions of the generalized ordering scheme given in Refs. [1] and [2], and this may be evident in the way the domain of integration is specified. Writing the resulting generalized ordered translational coordinates as $X_i'$, the domain of integration of Eqn. (1) may be specified such that the volume $v_i$ of the set of points traced out by $X_i'$ for the particle with label i has the property $0 \leq v_i \leq v_{i-1}$ (see Refs. [1] and [2] for a detailed discussion). Equation (1) may therefore be written as:

$$Z(N, V) = \int ... \int W_N dX_1' ... dX_N'$$

(3)

where $\sigma^N \widetilde{Z}(N, V) = Z(N, V)$. We may write

$$W_N(X_1', ..., X_N') = W_{N-1}(X_2', ..., X_N') \left[ \frac{W_N(X_1', ..., X_N')}{W_{N-1}(X_2', ..., X_N')} \right]$$

(4)

Taking $X_1'$ as origin and re-ordering the other generalized ordered coordinates according to distance from the origin, we may write $X_1$ as the coordinate (from among the remaining generalized ordered coordinates) which is nearest to the origin; $X_2$ the coordinate (from among the remaining generalized ordered coordinates) which is second nearest to the origin, etc. We may consider 2 clusters of coordinates; the first involving the coordinates $X_1', X_1, ..., X_n$ and the second involving $X_{n+1}, ..., X_{N-1}$. For some finite n which is large enough (but not expected to be very large), the second mild statement of the product property of the Slater Sum (given in section I) allows us write the term in square brackets in Eqn. (4) accurately as $\left[ \dfrac{W_{n+1}(X_1', X_1, ..., X_n)}{W_n(X_1, ..., X_n)} \right]$. (Accuracy in this case meaning that $|w_{n+1} - w_n| >> |f_{n+1} - f_n|$ or $f_{n+1} - f_n \sim 0$; see section I). Note that while the first cluster is of finite size, the second cluster is of infinite size in the thermodynamic limit N, V $\rightarrow \infty$ with N/V = $\rho$ the average number density. Eqn. (3) may therefore be written as:

$$Z(N, V) = \int ... \int Z(N-1, v_1) \left[ \frac{W_{n+1}(X_1', X_1, ..., X_n)}{W_n(X_1, ..., X_n)} \right] \times$$
$$\times \left[ W_{N-1}(X_2', ..., X_N') \Big/ \int ... \int W_{N-1}(Y_2, ..., Y_N) \prod_{i=2}^{N} dY_i \right] \prod_{j=1}^{N} dX_j'$$

$$(5)$$

where $Z(N-1, v_1) = \int ... \int W_{N-1}(X_2', ..., X_N') \prod_{i=2}^{N} dX_i'$. The volume within which the variables $X_2', ..., X_N'$ (as well as $Y_2, ..., Y_N$ which are also generalized ordered) are restricted, being $v_1$. The volume $v_1$ is that traced out by the coordinate $X_1'$ (c.f. Refs. [1] and [2]). From Eqn. (5), we may write:

$$Z(N,V) = \int Z(N-1,v_1)P(X_1',v_1)dX_1'$$

(6)

where

$$P(X_1',v_1) = \int \ldots \int \left( \frac{W_{n+1}}{W_n} \right) g^Q_{1,\ldots,N-1}(X_2',\ldots,X_N') \prod_{i=2}^{N} dX_i' \qquad (n \gg 1)$$

(7)

$g^Q_{1,\ldots,N-1}(X_2',\ldots,X_N')$ is the term in the second square brackets in the integrand of Eqn. (5). In the condition n >> 1, it is tacitly assumed that 'reasonably' accurate results can be obtained with n ≤ (a few tens).

The coordinates $X_i$ (i = 1, …, N-1) reordered according to distance from the origin $X_1'$ may also clearly be said to be ordered according to one of the versions of the generalized ordering scheme.[1,2] If for instance the first version [1,2] of the generalized ordering scheme is employed and the coordinates $X_i$ (i = 1, …, N-1) are restricted within the volume $v_1$, then the complete coordinate set $X_1', X_1, \ldots, X_{N-1}$ may be said to constitute a set which is ordered according to the first version of the generalized ordering scheme. Hence this new set of coordinates may be used to replace the set $X_1',\ldots,X_N'$ in Eqns. (3) to (7). The function $g^Q_{1,\ldots,N-1}(X_1,\ldots,X_{N-1})$ may therefore be seen as the analogue of the NNPDF for classical systems of Ref. 1 and is referred to as the Quantum Nearest Neighbor Probability Density Function (QNNPDF). Because $W_N$ is proportional to an actual probability density function in coordinate space, QNNPDF's are also actual probability density functions in coordinate space for quantum systems at least in the non-relativistic limit.

Integrating $g^Q_{1,\ldots,N-1}(X_1,\ldots,X_{N-1})$ over all allowable values of the coordinates $X_i$ (i = n+1, …, N-1) yields the "general point process" [5] QNNPDF for n nearest neighbors (with origin at a point on the boundary of volume $v_1$) and written as $g^Q_{1,\ldots,n}(X_1,\ldots,X_n)$. Generalizations to the "ordinary point process" [5] and cases where the origin is situated in the middle of the volume V are rather obvious. QNNPDF's will be further discussed in the next section.

Equation (7) may be rewritten as:

$$P(X_1',v_1) = \int \ldots \int \left( \frac{W_{n+1}(X_1',X_1,\ldots,X_n)}{W_n(X_1,\ldots,X_n)} \right) g^Q_{1,\ldots,n}(X_1,\ldots,X_n) \prod_{i=1}^{n} dX_i \qquad (n \gg 1)$$

(8)

$W_n$ features in the function $P(X_1',v_1)$ as well as in $g^Q_{1,\ldots,n}(X_1,\ldots,X_n)$ (see next section), and thus it is necessary to have an accurate and efficient way of computing $W_n$ for various nearest neighbor configurations. In the semi-classical approach to statistical mechanics, various approximate methods [4] exist for computing $W_n$ analytically. However, we expect that other means, such as the techniques of Quantum Monte Carlo (MC) methods [6], which hitherto may not have been directly used to compute $W_n$, should be more suitable especially for handling the Boltzmann's factor in $W_n$, in order to yield accurate results for $W_n$. Writing $W_n$ (the diagonal element of the Boltzmann's operator in coordinate representation) as $n!\lambda^{3n}\langle X_1,\ldots,X_n | e^{-\beta H_n} | X_1,\ldots,X_n \rangle$, we can recast this expression in various forms (such as by inserting the identity operator), and thus expect

to ultimately carry out computations with the Boltzmann's operator in a desired representation. This avoids for instance, the use of operators in differential form, for the Hamiltonian $H_n$, in the Boltzmann's operator. Thus allowing for a considerably more efficient way to compute $W_n$. The multiple integral for $P(X_1', v_1)$ may also be efficiently computed by numerical methods for instance, as described for the pure classical case.[1] Unlike usual Quantum MC works, the computation of $P(X_1', v_1)$ is not burdened with the requirement that n must ideally be indefinitely large. Instead, our n is required not only to be finite (by the argument on surface-to-volume ratio earlier provided in section I), but also, n is required to be "small" or not too large (owing to the logarithmic nature of the mild statement on the product property of the Slater Sum).

Extreme difficulties are known to exist when attempting to determine "reasonable" results for such quantities as $W_n$. However, it is in cases of "small" n values one may expect to achieve considerable progress in the bid to obtain accurate results. Hence it is not far fetched to expect that highly efficient methods should soon be available for accurately evaluating such quantities as $W_n$, $g^Q_{1,\dots,n}(X_1,\dots,X_n)$, and $P(X_1', v_1)$ for small n. However the case may be, it is clear that one of the major achievements of this paper is the considerable reduction of the degree of difficulty of the original quantum problem without compromising the accuracy of desired results.

The above discussion briefly outlines the numerical computational aspects of our scheme. This numerical aspect is definitely not nearly as computationally intensive as usual MC methods (which technically are required to investigate systems in the limit n $\rightarrow$ $\infty$). Thus our scheme falls into a new category of analytical methods whose numerical aspects have computational intensity which fall between those of usual Monte Carlo work, and those of purely analytical methods. (A feature which becomes very interesting if considered as a new "requirement standard" for the efficient and accurate investigation of many-body problems).

Because $P(X_1', v_1)$ may generally vary as $X_1'$ varies for fixed $v_1$, we may therefore replace P by $\langle P \rangle$ as in the classical case [1,2] and thus rewrite Eqn. (6) as:

$$Z(N,V) = \int_0^V Z(N-1, v_1) \langle P \rangle dv_1$$

(9)

As in Refs. [1] and [2], we expect $\langle P \rangle \approx P$ almost always in many cases of interest. Clearly, $W_N$ is dimensionless, hence we may write:

$$Z(N,V) = \frac{(\varepsilon V)^N}{N!}$$

(10)

where ε is a dimensionless quantity which is some function of N, V. Note that in the non-interaction limit, ε is not expected to tend to unity in general, as in a poisson point process [5] (owing to quantum effects), suggesting the quantum case may generally be considerably more computationally involved than the classical case. However, we expect $\varepsilon \to 1$ as $\rho \to 0$. Proceeding as in Ref. [1], we obtain the equation:

$$\varepsilon \left(1 - \frac{\rho}{\varepsilon} \frac{\partial \varepsilon}{\partial \rho}\right) = \langle P \rangle \exp\left(-\frac{\rho}{\varepsilon} \frac{\partial \varepsilon}{\partial \rho}\right)$$

(11)

Equation (11) is then readily solved via an iterative scheme for ε employing methods outlined in Ref. [1]. In the next section, we show that QNNPDF's depend on ε; hence as ε gets more accurately determined by the iterative process, QNNPDF's also get more accurately evaluated. At the end of the iterative process, the free energy (obtained from ε) or QNNPDF's may be employed to accurately compute a variety of system properties.

If the "effective" pair-wise inter-particle interaction potential is largely weak and slowly varying (so that at worst, its characteristic distance is $>> \lambda$), and if the average distance between the coordinates of the n nearest neighbors ($X_1, \ldots, X_n$) is also $>> \lambda$, we have that $W_n$ approximates to[4] $\left( \dfrac{\lambda}{h} \right)^{3n} \int \ldots \int e^{-\beta H_n} dP_1 \ldots dP_n = e^{-\beta U_n}$ . ($H_n$ is the classical Hamiltonian, and $P_i$ is the momentum conjugate to $X_i$). Hence $\left( \dfrac{W_{n+1}}{W_n} \right)$ reduces to the quantity $e^{-\beta E_2}$ of Ref. [1], and the quantum calculations for Z(N, V) coincides with those of the classical case. This coincidence clearly occurs at low densities or high temperatures where the QNNPDF is expected to give much more weight to configurations in which the coordinates are far apart compared to $\lambda$.

# III. NNPDF's IN

# QUANTUM SYSTEMS

In the previous section QNNPDF's were briefly introduced and shown to be the counterparts of (classical case) NNPDF's for describing structure in quantum systems. We now develop them further in the present section. An origin may be defined somewhere in the middle of the volume V. The coordinates are assumed ordered according to distance from the origin. ie., $r_{i+1} \geq r_i$ for i = 1, …, N-1. ($r_i$ is the radial part of $X_i$). By this, it is clear the coordinates are also generalized ordered. From earlier discussions, we may write the general point process QNNPDF for all N coordinates as:

$$g^{QG}_{1,...,N}(X_1,...,X_N) = \exp\left(\frac{A}{KT}\right)W_N(X_1,...,X_N)$$

("A" is free energy of the system). Writing $W_N$ as $W_{N-n}\ W_n\ F_n$ and integrating the above expression over all coordinates except the first n coordinates, we get the QNNPDF for n-nearest neighbors in the general point process as:

$$g^{QG}_{1,...,n}(X_1,...,X_n) = \exp\left(\frac{A}{KT}\right)W_n\int...\int W_{N-n}F_n\prod_{i=n+1}^{N}dX_i$$

(12)

The coordinates $X_i$ (i = n + 1, …, N) are restricted within the volume $\hat{V} = V - \frac{4}{3}\pi r_n^3$ which has a different "shape" than that of volume V. Following the method outlined in section III of Ref. 5, Eqn. (12) is readily manipulated to arrive at:

$$g^{QG}_{1,\dots,n}(X_1,\dots,X_n) = \left[\exp\left(\frac{A}{KT}\right)Z(N-n,V)\right]\left[\frac{^{\hat{V}}S_{N-n}\langle F_n\rangle Z(N-n,\hat{V})}{Z(N-n,V)}\right][W_n]$$

(13)

$^{\hat{V}}S_{N-n}$ is a shape effect factor which accounts for the difference in shape between the volume $\hat{V}$ which has a "void" of size $\frac{4}{3}\pi r_n^3$ (containing the origin and the n-nearest neighbors) located within it, and some volume of "standard" shape (without a void) of the same size. (Note that the shape of volume V actually defines the standard shape – see Ref. [5] for a detailed discussion). $\langle F_n\rangle$ is an average taken over the phase space of size $Z_s(N-n,\hat{V})=^{\hat{V}}S_{N-n}Z(N-n,\hat{V})$. (The quantity $Z(N-n,\hat{V})$ is the partition function evaluated using the volume of standard shape of size $\hat{V}$). $^{\hat{V}}S_{N-n}$ and $\langle F_n\rangle$ provide shape and surface effects respectively as discussed in Ref. [5]. Expressing Z(N-n, V) in terms of ε and noting |f_n|, << |w_n,|, |w_{N-n}| for n sufficiently large (c.f. section I), we may argue as in Ref. [5] that because surface and shape effects present themselves in "reduced" forms, they may be ignored (similar to small computational errors) once we have chosen n to be at least as large as some value that is "small" (probably a few tens or so). Hence Eqn. (13) may be rewritten as:

$$g^{QG}_{1,\dots,n}(X_1,\dots,X_n) = h_n \exp\left[ -\frac{4}{3}\pi r_n^{\,3}\rho\left(1 - \frac{\rho}{\varepsilon}\frac{\partial\varepsilon}{\partial\rho}\right)\right] W_n(X_1,\dots,X_n)$$

$$= h_n \exp\left[ -\frac{4}{3}\pi r_n^{\,3}\, {p}\middle/{KT}\right] W_n(X_1,\dots,X_n)$$

$$(n \gg 1) \tag{14}$$

$h_n$ is a normalization constant, and p is the system's pressure. Use has been made of the fact that we can write the equation of sate as [1]:

$$\phi = \frac{p}{KT} = \rho\left(1 - \frac{\rho}{\varepsilon}\frac{\partial\varepsilon}{\partial\rho}\right)$$

$$\tag{15}$$

Eqn. (14) is "exact", and it rigorously describes structure in quantum systems. In the "ordinary" point process [5], we replace the quantity N by (N − 1). Also, $W_n(X_1, \dots, X_n)$ is rewritten as $W_{n+1}(X_0, X_1, \dots, X_n)$ where $X_0$ is the origin known to contain the original particle. The quantity "A" becomes the free energy of (N − 1) particles with external influence from the original particle placed at the origin.

For a general point process in which the origin is located at a point on the boundary surface of volume V, it is easy to see that Eqn. (14) becomes:

$$g^{Q}_{1,\dots,n}(X_1,\dots,X_n) = h_n \exp\left( -\frac{2}{3}\pi r_n^{\,3}\, {p}\middle/{KT}\right) W_n(X_1,\dots,X_n) \qquad (n \gg 1)$$

$$\tag{16}$$

Note here that the boundary surface of volume V is assumed locally flat everywhere (hence the factor of $\frac{4}{3}$ in Eqn. (14) simply changes to $\frac{2}{3}$ in the above equation in the thermodynamic limit (c.f. Refs. [1] and [5]). In the ordinary point process in this case, we simply replace $W_n$ by $W_{n+1}$ as previously explained. Assuming $v_1$ is large enough "almost always" (this is usually the case), we find Eqn. (16) is what is required to be employed for the QNNPDF in Eqn. (8). Note in this case however, that the effect of the particle said to be located at the origin $X_1'$ which is a point on the boundary surface of volume $v_1$ is not taken into account in the computation of $W_n(X_1, \ldots, X_n)$. At low temperatures and/or high densities, we find that except for small $r_n$ values, the argument of the exponential term in Eqn. (16) is generally a large negative number and this clearly has the effect of making the average of $r_n$ to be small. (This is a "push" effect from without [5], and this effect generally predominates at low temperatures and/or high densities). Hence more weight is given under this condition to configurations in which the n-nearest neighbors are "relatively" closely spaced apart. In the limit of low density and/or high temperatures, it becomes obvious that considerably more weight now shifts to configurations where coordinates are "relatively" far apart as earlier noted. QNNPDF's may be said to complement quantum n-body distribution (or correlation) functions [3,4,5] for describing structure in quantum systems. In subsequent work, we will relate QNNPDF's to quantum n-body distribution functions as was done for the classical case [5].

In the low density and/or high temperature limit, $W_n$ behaves as $\sim e^{-\beta U_n}$ (as earlier indicated) and hence in this limit, QNNPDF's are seen to behave as classical NNPDF's. Inclusion of external forces in the evaluation of the free energy and structure of quantum systems may also be investigated exactly, following the counterpart approach outlined for the classical case. [1] In a later work, we anticipate providing similar developments as above for quantum mixed systems as has been recently done for classical multi-component systems.[2] Structure in this case will be expected to involve what may be termed partial QNNPDF's (or **PQNNPDF's**).

# IV. APPLICATION TO WEAKLY DEGENERATE QUANTUM SYSTEMS

For ease of computation, the scheme developed in sections II and III shall now be applied to simple quantum systems that are weakly degenerate, where the average inter-particle spacing is much larger than the thermal wavelength. It was already noted for weakly degenerate systems that $W_n \sim e^{-\beta U_n}$ implying that the NNPDF's discussed in the previous section revert to those of the classical system. For the ideal quantum fluid that is weakly degenerate, the corresponding NNPDF's become those of the Poisson fluid [5]. The Poisson result also applies for the weakly degenerate **non**-ideal quantum system with sufficiently low density (where the average inter-particle spacing is considerably larger than the range of effective pair-wise interaction potential).

To determine the free energy, we begin by substituting Eqn. (16) into Eqn. (8) to obtain:

$$P(X_1, v_1) = \int \ldots\ldots\ldots\int_{(half-space)} h_n \exp\left(-\frac{2}{3}\pi r_n{}^3 \middle/ \frac{P}{KT}\right) W_{n+1}(X_1', X_1, \ldots, X_n) \prod_{i=1}^{n} dX_i$$

$h_n$ is obtained by normalizing $g^Q{}_{1,\ldots,n}(X_1, \ldots, X_n)$. For the generally non-ideal quantum system which is weakly degenerate (with sufficiently low density), the NNPDF approximates to that of the Poisson fluid. This implies we may accurately write[5] $h_n = \rho^n$.

In the low density limit, surface and shape effects are expected to be weak, and NNPDF's as formulated in the previous section may be expected to yield accurate results for small n (say n = 1). Also, since at sufficiently low densities a particle may be said to interact mainly with its nearest neighbor, accurate results are expected using n = 1. In which case, P may be well approximated as:

$$P(X_1', v_1) \cong \int\limits_{(half-space)} \rho \exp\left(-\frac{2}{3}\pi r_1^3 \, {}^P\!/_{KT}\right) W_2(X_1', X_1) dX_1$$

(17)

But in the absence of external forces, $W_2(X_1', X_1)$ depends only on the distance between the original particle situated at $X_1'$ and its nearest neighbor (see Ref. [4]). Eqn. (17) may be rewritten as:

$$P(X_1', v_1) \approx \int\limits_0^\pi \int\limits_0^\pi \int\limits_0^\infty \rho r_1^2 \exp\left(-\frac{2}{3}\pi r_1^3 \rho\right) W_2(r_1) \sin\theta_1 dr_1 d\theta_1 d\varphi_1 = 2\pi\rho \int\limits_0^\infty r_1^2 \exp\left(-\frac{2}{3}\pi r_1^3\right) W_2(r_1) dr_1$$

(18)

At sufficiently low densities, Eqn. (11) shows that ε ≈ P (keeping terms up to first order power in ρ). Hence by integrating Eqn. (18) by parts, we may write:

$$\varepsilon \approx -\exp\left(-\frac{2}{3}\pi\rho r_1^3\right) W_2(r_1)\Big|_{r_1=0}^{r_1=\infty} + \int\limits_0^\infty \exp\left(-\frac{2}{3}\pi\rho r_1^3\right) W_2'(r_1) dr_1$$

where $W_2'(r_1) = \dfrac{dW_2(r_1)}{dr_1}$. This then yields:

$$\frac{d\varepsilon}{d\rho} = \int\limits_0^\infty \left(-\frac{2}{3}\pi r_1^3\right)\exp\left(-\frac{2}{3}\pi\rho r_1^3\right)W_2'(r_1)dr_1$$

(19)

We may assume the integral $\int\limits_0^\infty r_1^3 W_2'(r_1)dr_1$ not only exists, but also has a finite range

$(0 \le r_1 \le R$ say) over which it has its major contribution. In which case, if the density is so low that we may approximate the exponential term in Eqn. (19) as $\approx 1$ for $0 \le r_1 \le R$, we then write in the low density limit:

$$\frac{d\varepsilon}{d\rho} = -\int\limits_0^\infty \frac{2}{3}\pi r_1^3 W_2'(r_1)dr_1$$

(20)

In the weakly degenerate state, $W_N$ was said to be appropriately approximated as $e^{-\beta U_N}$; hence the partition function becomes that of the classical case, and in the low density limit, we may approximately write $\varepsilon \approx 1$. The equation of state (Eqn. (15)) in the low density limit may therefore be written as:

$$\phi\Big/\rho = 1 + \rho\int\limits_0^\infty \frac{2}{3}\pi r_1^3 W_2'(r_1)dr_1$$

(21)

But the integral term of Eqn. (21) is readily shown to be the second virial coefficient for quantum systems. The second virial coefficient is given in the literature as:

$$-\frac{1}{2}\int\left(W_2(r_1)-1\right)d^3r = -\int_0^\infty 2\pi r^2\left(W_2(r)-1\right)dr$$

(see for instance the paper of Kahn and Uhlenbeck[4]). Integrating this by parts yields the integral term of Eqn. (21). (Note that the result[4] $\lim_{r\to\infty} W_2(r) \to 1$ has been employed).

In moderately/highly degenerate quantum systems where there has been extreme difficulty of investigation, the method of this paper is expected to require extensive local neighborhood MC calculations with n finite but relatively small (n ~ a few tens). This is clearly because it rapidly becomes a very poor approximation to assume a particle may interact only with its first nearest neighbor.

# V. ADDITIONAL COMMENTS AND CONCLUSIONS

The accomplishments of the methods expounded in Refs. [1], [2], and [5] for classical systems have been duplicated here for quantum systems. The extended notion of NNPDF's (ie. QNNPDF's), along with the notions of generalized order, and the one particle phase space have once again been employed, this time to obtain a new formalism for investigating the statistical thermodynamics of quantum systems. Application of the above formalism to the weakly degenerate quantum system was demonstrated. Just as in the classical case, it is expected that the degree of success achieved will not be abated, as the present scheme is progressively applied to more complex systems (strongly correlated/interacting many-body systems).

The coordinate representation employed in this paper in expressing the Boltzmann's operator, is known to be just as valid (from a purely mathematical stand point) as other preferred representations in quantum developments used to determine values for "observable quantities". Thus the method of the present paper is readily seen to have general applicability. Another important notion deals with the number of particles. Hence in some situations a grand-canonical ensemble becomes an important extension of the scheme of this paper. It is interesting to observe that the underlying formalism of this paper was actually developed for all allowed particle number N (see Eqn. 9 of Ref. [1] or Eqn. 8 of Ref. [2], which lead in general to difference equations rather than differential equations). Hence other asymptotic expressions may be developed for various quantities of the present scheme for different regimes of N, and such expressions may then be employed in determining a grand ensemble formalism.

Our approach essentially reduces a many-body problem (n → ∞) to a "few-body" problem (n ~ a few tens), whereby, the most computationally intensive aspect of the scheme employs numerical (or MC) calculations for a subsystem involving a few nearest neighbor particles to determine the statistical parameter $\langle P \rangle$. The scheme then provides an "extrapolation" of the results for $\langle P \rangle$ for the finite subsystem to results for the infinite system. In which case, results for $W_n$, and $\langle P \rangle$ (which are results for a **finite particle cluster**), may be said to yield a "signature" from which properties of the infinite (many-body) system are inferred. This is very interesting as current literature[7] attests to the fact that small clusters of particles sometimes do possess certain properties of the infinite system. It is not unexpected however, that these signatures generally involve properties of small particle clusters, which may sometimes differ considerably, or sometimes differ in subtle ways, from those of the infinite system.

It is known in the literature that extreme difficulties exist in determining quantities such as $W_n$ (for all n) in quantum systems. A major point of this paper however, is that, the method developed greatly reduces the degree of complexity of the quantum statistical thermodynamic problem without compromising the accuracy of results obtained for the original problem. This points at the fact that attempts to determine accurate results for $W_n$ are not only worthwhile, but may be achievable at least in cases of small n values.

We note that the product property for $W_N$ (which plays the role of the Boltzmann's factor for classical systems) was instrumental in developing a strong analogy between the scheme of this paper, and that used for classical systems.[1] It is interesting also to note that the use of the product property allowed us avert the sign problem usually encountered in fermion calculations. It is our intention to continue to explore the product property beyond what may be presently available in the literature. We recognize however, that quantum MC methods may be indispensable in this regard in consolidating findings obtained by purely analytical means and especially in establishing the minimum size of n necessary for the "few-body" calculations mentioned earlier. Aspects of the current scheme that need urgent development, if the method of this paper may be used for realistic 3D material systems, are the numerical calculations for $W_n$ and <P>. Hence the quantum MC scheme for this kind of numerical computations is amongst the set of investigations currently considered of top-most priority.

# ACKNOWLEDGEMENTS

This work was supported in part by the University of Wisconsin – Madison.